\documentclass[journal]{ieeetran}
\usepackage{cite}
\usepackage{wrapfig}
\usepackage{amsmath,amsfonts,amssymb,bm}
\usepackage{mathtools,nccmath}
\usepackage{algorithmic}
\usepackage{graphicx}
\usepackage{textcomp}
\usepackage[ruled,vlined]{algorithm2e}
\usepackage{booktabs}
\usepackage[utf8x]{inputenc}
\usepackage{diagbox}
\usepackage{multirow}
\usepackage{arydshln}
\usepackage{dblfloatfix}
\usepackage{url}

\begin{document}
\title{	Outlier-Robust Filtering For Nonlinear Systems With Selective Observations Rejection}

\author{Aamir Hussain Chughtai, Muhammad Tahir, \IEEEmembership{Senior Member, IEEE}, and Momin Uppal, \IEEEmembership{Senior Member, IEEE}
	\thanks{The authors are with Department of Electrical Engineering, Lahore University of Management Sciences, DHA Lahore Cantt., 54792, Lahore Pakistan. { (email: aamir.chughtai@lums.edu.pk; tahir@lums.edu.pk; momin.uppal@lums.edu.pk)}}%
}

\maketitle

\begin{abstract}
	Considering a common case where measurements are obtained from independent sensors, we present a novel outlier-robust filter for nonlinear dynamical systems in this work. The proposed method is devised by modifying the measurement model and subsequently using the theory of Variational Bayes and general Gaussian filtering. We treat the measurement outliers independently for independent observations leading to selective rejection of the corrupted data during inference. By carrying out simulations for variable number of sensors we verify that an implementation of the proposed filter is computationally more efficient as compared to the proposed modifications of similar baseline methods still yielding similar estimation quality. In addition, experimentation results for various real-time indoor localization scenarios using Ultra-wide Band (UWB) sensors demonstrate the practical utility of the proposed method.  
\end{abstract}

\begin{IEEEkeywords}
	Sensor Data Outliers, Sensor Degradation, Online Robust Filtering, General Gaussian Filtering, Unscented Kalman Filter, Variational Bayes, Approximate Inference, Indoor localization, Ultra-wide band (UWB) sensors.
\end{IEEEkeywords}

\maketitle

\section{Introduction}
Sensors play an important role in the functionality of different physical dynamical systems providing useful data for estimation of key system parameters and effectively controlling the underlying processes. To describe dynamical systems, State-Space Models (SSMs) are widely used in diverse applications such as cyberphysical systems,  robotics, sensor fusion, navigation, guidance, and tracking systems \cite{ghysels2018applied,doi:10.1162/089976603765202622,1100844,7435151,paninski2010new, 9447033,4154676,7001546}. In an SSM, the system is described by latent states evolving with first-order (Markovian) dynamics. The states are not directly observable rather manifest through a set of external outputs measured by sensors.

Kalman Filter (KF) and its variants \cite{kalman1960new,wishner1969comparison,wan2001unscented} are benchmark state estimation methods for SSMs. These filters assume precise apriori knowledge of the system's noise statistics. However, in practice, the data from different sensors can easily be corrupted with outliers due to factors like inherent sensor quality or its degradation over time, communication glitches, environmental effects etc. This results in divergence of the actual and assumed noise statistics making KFs ineffective\cite{9239326}. Therefore, the development of robust filters remains a busy research area \cite{9447033, 8891731, 9113473}.


Several popular robust filtering methods, classified as compensation-based methods, aim to utilize information from the outlier-ridden observations for inference updates. Some of these techniques assume prior statistics of the measurement noise or the residuals. These include methods based on robust statistics \cite{karlgaard2015nonlinear,7855662,chang2012multiple} and methods based on modeling the observation noise as Student-t or Laplacian distribution \cite{7812899,8009803}. Their performance is effective owing to the sporadic nature of outliers. However, since these methods are based on static loss functions based on design parameters, meticulous tuning of these parameters is required \cite{8869835}. Therefore, tuning-free learning-based compensation approaches have also been advocated in the literature \cite{6657710,du2015observation,6349794,8869835}. These methods assume a distribution to describe the measurement noise and subsequently aim to learn the parameters of the distribution.

Other methods, known as rejection-based approaches, stem from the argument that generally outliers come from clutter and do not necessarily obey a well-defined
distribution. Therefore, corrupted measurements should be completely discarded for state estimation. Traditionally, this is performed by comparing the normalized measurement residuals with some predefined thresholds \cite{6112697,6616007,8245799}. However, the selection of the threshold is mostly subjective. Some theoretical justifications for threshold selection are provided in an extended KF (EKF) based method \cite{mu2015novel}. However, the method is tested only for low outliers frequency and needs memory for past observations. Different learning-based rejection approaches have also been proposed in literature \cite{8398426,9239326}. These techniques aim to learn the parameters that determine whether to use (or discard) the observations for state estimation.

Compared to the traditional counterparts, learning-based robust filters have advantages in terms of minimizing user discretion, being more general, and better suited for single-shot applications \cite{8869835}. As exact inference gets analytically infeasible in their development, approximate inference techniques like Particle Filters (PFs) and Variational Bayesian (VB) methods are usually employed. Since PFs can be computationally very expensive, VB methods are the popular alternatives to devise tractable robust filters. With this background, we only consider VB-based outlier-robust filters in this work. Note that for robust estimation non-Bayesian paradigms have also been investigated e.g. considering ambiguity sets to cater for model distributional uncertainties, formulating and solving different optimization problems etc.\cite{shafieezadeh2018wasserstein,tzoumas2019outlier,8047245}. Connections between the Bayesian and other approaches are being currently explored under different assumptions \cite{shafieezadeh2018wasserstein,nguyen2019bridging}. In this work, we restrict our attention to the Bayesian methods for robust filtering.

In several applications, the measurements are commonly obtained from independent sensors e.g. in applications like robotics, wireless sensor networks,  Internet of Things (IoT) etc. As the number of sensors and data acquisition rates can be high in such applications, with limitations on the processing power, specialized outlier-robust filtering is required. Estimation quality and computational overhead are the primary design criteria for devising such filters. A review of existing works indicates that tractable learning-based robust filters are generally under-parameterized i.e. devised
considering common parameters for all the measurement
dimensions for outlier compensation. While it is acceptable when the measurements are inter-dependent, it results in loss of useful information and hence robustness, for independent observations. 

\subsubsection*{Contributions} 
The specific contributions of this work are as follows.
\begin{itemize}
\item We offer a better parameterization of outliers to avoid needless information loss. We treat the outliers independently when the measurements are obtained from independent sensors.
\item Using VB inference and general Gaussian filtering, we devise an outlier-robust learning-based tuning-free filter for nonlinear SSMs, which discards only the corrupted measurements during inference. We consider nonlinear systems in our study in contrast to the methods that are devised for linear systems \cite{chen2017maximum,huangnovel}. Unlike the method in \cite{9050910}, we do not take any restrictive assumption that outliers can occur in at most one measurement dimension at any given instant. 
\item In addition, we propose modifications to the existing baseline methods to enable them deal with outliers selectively for each dimension. 
\item We also present a modification to our originally devised method that is computationally more efficient as the modifications of the rival methods and exhibit similar estimation quality.
\item By comparing our method with other learning-based tractable approaches using simulations we verify the performance gains. Moreover, experimental evaluation for different indoor localization scenarios using UWB sensors shows the practical efficacy of the proposed filter.
\end{itemize}

We present the remaining article with the following organization. In Section \ref{Sec_model} modeling description of the proposed filter is presented. Section \ref{Sec_VB_infer} provides derivation details of the basic proposed method. Subsequently, in Section \ref{Eval} performance evaluation results of the considered methods and their proposed modifications are discussed. Finally, the conclusions drawn from the work are reported in Section \ref{Concl}.

\section{Modeling Details}\label{Sec_model}
We consider a nonlinear discrete-time SSM of a dynamic physical process mathematically described by
\begin{align}
\mathbf{x}_k&= \mathbf{f}(\mathbf{x}_{k-1})+\mathbf{q}_{k-1}	
\label{eqn_model_1}\\
\mathbf{y}_k&= \mathbf{h}(\mathbf{x}_{k})+\mathbf{r}_{k}
\label{eqn_model_2}
\end{align}
where the subscript $k$ denotes the time index; $\mathbf{x}_k\in \mathbb{R}^n$ and $\textbf{y}_k \in \mathbb{R}^m$ are the state and measurement vectors respectively; the non-linear functions  $\textbf{f}(.):\mathbb{R}^n\rightarrow\mathbb{R}^n$ and $\textbf{h}(.):\mathbb{R}^n\rightarrow\mathbb{R}^m$  define the process dynamics and observation equations respectively; $\textbf{q}_{k}\in \mathbb{R}^n$ and $\textbf{r}_k\in \mathbb{R}^m$ account for the process and measurement noise respectively. $\textbf{q}_{k}$ and $\textbf{r}_k$ are assumed to be statistically independent, White and normally distributed with zero mean and predetermined covariance matrices $\textbf{Q}_k$ and $\textbf{R}_k$ respectively. In addition, as a general notation in this work, ${{a}^i}$ denotes the $i$th element of a vector $\textbf{a}$, ${\mathrm{A}^{ii}}$ denotes the $i$th diagonal element of a matrix $\mathbf{A}$ and ${diag}(\textbf{a})$ is a diagonal matrix with the elements of $\textbf{a}$ as its entries.

Practically, the observations from different sensors can be corrupted with outliers. This leads to the failure of conventional filtering based on the model \eqref{eqn_model_1}-\eqref{eqn_model_2}. In the following, we consider that observations are obtained from independent sensors, therefore, we model the outliers independently for each dimension. To mitigate the effect of outliers on the state estimation quality, we introduce an indicator vector $\bm{\mathcal{I}}_k\in\mathbb{R}^m$ with Bernoulli elements. In particular, ${{\mathcal{I}}}^i_k$ can assume two possible values $\epsilon$ (close to zero) and 1.  ${{\mathcal{I}}}^i_k=\epsilon$ indicates the occurrence of an outlier in the corresponding dimension at time $k$. Since an outlier can occur independently at any instant, irrespective of the past and outliers in other dimensions, we assume that the elements of $\bm{{\mathcal{I}}}_k$ are statistically independent of each other and their history. Moreover, $\bm{\mathcal{I}}_k$ and $\mathbf{x}_k$ are also considered independent. Using $\theta^i_k$ to denote the probability of no outlier in the $i$th observation, the distribution of $\bm{\mathcal{I}}_k$ is expressed as

\begin{equation}
p(\bm{\mathcal{I}}_k)=\prod_{i=1}^{m}p({{\mathcal{I}}}^i_k)=\prod_{i=1}^{m} (1-{\theta^i_k}) \delta({{{\mathcal{I}}}^i_k}-\epsilon)+{\theta^i_k}\delta( {{{\mathcal{I}}}^i_k}-1)
\label{eqn_model_3}
\end{equation}

Furthermore, the measurement likelihood conditioned on the current state $\mathbf{x}_k$ and the indicator $\bm{\mathcal{I}}_k$, independent of all the historical observations $\mathbf{y}_{1:{k-1}}$, is proposed to follow a Gaussian distribution

\begin{align}
&p(\mathbf{y}_k|\mathbf{x}_k,\bm{\mathcal{I}}_k)={\mathcal{N}}\Big(\mathbf{y}_k|\mathbf{h}(\mathbf{x}_k),\bm{\Sigma}_k^{-1}\Big)\nonumber\\
&=\frac{1}{\sqrt{(2 \pi)^{m}|\bm{\Sigma}_k^{-1}|}}\mathrm{exp}\big\{{-}\mfrac{1}{2}{(\mathbf{y}_k-\mathbf{h}(\mathbf{x}_k))}^{\text{T}}\bm{\Sigma}_k(\mathbf{y}_k-\mathbf{h}(\mathbf{x}_k))\big\}	
\label{eqn_model_4}\\
&=\prod_{i=1}^{m}\frac{1}{\sqrt{2 \pi {\mathrm{R}^{ii}_k}/{{\mathcal{I}}}^i_k}}\ \mathrm{exp}\big\{{-}\frac{{(\mathrm{y}^i_k-\mathrm{h}^i(\mathbf{x}_k))}^{2}}{2\mathrm{R}^{ii}_k}{{\mathcal{I}}}^i_k\big\}
\end{align}
where $\bm{\Sigma}_k={\textbf{R}^{-1}_k}{diag}(\bm{\mathcal{I}}_k)$.  We assume statistical independence in the nominal noise adding in each of the measurement dimension as commonly observed especially in cases where the sensors are deployed independently. Therefore, $\textbf{R}_k$ is assumed to be diagonal and we are able to express the distribution as a product of univariate Gaussian distributions.

{Considering \eqref{eqn_model_3} and \eqref{eqn_model_4}, the modified measurement model incorporating the effect of outliers can be expressed as
\begin{align}
\mathbf{y}_k&= \mathbf{h}(\mathbf{x}_{k})+\bm{\nu}_{k}
\label{eqn_model_2-m}
\end{align}
where the modified measurement noise assumes a Gaussian mixture model as $\bm{\nu}_{k}\sim \underset{{\bm{\mathcal{I}}_k}}{\sum} \ {\mathcal{N}}(\bm{\nu}_k|\mathbf{0},\bm{\Sigma}_k^{-1}) p(\bm{\mathcal{I}}_k)$.}

\section{Variational Bayesian Inference}\label{Sec_VB_infer}
With the proposed observation model, we can employ the Bayes rule recursively to obtain the analytical expression of the joint posterior distribution of $\mathbf{x}_k$ and $\bm{\mathcal{I}}_k$ conditioned on the set of all the observations $\mathbf{y}_{1:{k}}$
\begin{equation}
p(\mathbf{x}_k,\bm{\mathcal{I}}_k|\mathbf{y}_{1:{k}})=\frac{p(\mathbf{y}_k|\bm{\mathcal{I}}_k,\mathbf{x}_{k})	p(\mathbf{x}_k|\mathbf{y}_{1:{k-1}})p(\bm{\mathcal{I}}_k)}{p(\mathbf{y}_k|\mathbf{y}_{1:{k-1}})}
\label{eqn_vb_1}
\end{equation} 
The joint posterior can further be marginalized to obtain $p(\mathbf{x}_k|\mathbf{y}_{1:{k}})$ for state inference. However, using this approach directly is computationally complex. Therefore, we resort to the standard VB method \cite{vsmidl2006variational}, {a technique for approximating intractable integrals arising in Bayesian inference}, where the joint posterior is approximated as a product of marginal distributions 
\begin{equation}
p(\mathbf{x}_k,\bm{\mathcal{I}}_k|\mathbf{y}_{1:{k}})\approx q(\mathbf{x}_k)q(\bm{\mathcal{I}}_k)
\label{eqn_vb_2}
\end{equation} 
The VB approximation aims to minimize the Kullback-Leibler (KL) divergence between the product approximation and the true posterior. Accordingly, with $ \langle.\rangle_{q(\bm{\psi}_k)}$ denoting the expectation of the argument with respect to a distribution $q(\bm{\psi}_k)$, the variational distributions can be updated in an alternating manner as   
\begin{align}
q(\mathbf{x}_k)&\propto \mathrm{exp}\big( \big\langle\mathrm{ln}(p(\mathbf{x}_k,\bm{\mathcal{I}}_k|\mathbf{y}_{1:{k}})\big\rangle_{q(\bm{\mathcal{I}}_k)}\big)\label{eqn_vb_3}\\
q(\bm{\mathcal{I}}_k)&\propto\mathrm{exp}\big( \big\langle\mathrm{ln}(p(\mathbf{x}_k,\bm{\mathcal{I}}_k|\mathbf{y}_{1:{k}})\big\rangle_{q(\mathbf{x}_k)}\big)\label{eqn_vb_4}
\end{align} 
For tractability, we integrate general Gaussian filtering \cite{sarkka2013bayesian} results into the VB framework by assuming $p(\mathbf{x}_k|\mathbf{y}_{1:{k-1}})\approx {\mathcal{N}}\left(\mathbf{x}_k|\mathbf{m}^{-}_k,\mathbf{P}^{-}_{k}\right)$. Using the expressions of the prior distributions and measurement likelihood in \eqref{eqn_vb_1}, the posterior distribution is approximated as a product of marginals derived as follows 
\subsection{Derivation of $q(\mathbf{x}_k)$}
Using \eqref{eqn_vb_3} we can write
\begin{align}
q(\mathbf{x}_k)\propto \mathrm{exp}&\big\{-\mfrac{1}{2} (\mathbf{y}_k-\mathbf{h}(\mathbf{x}_k))^{\text{T}}\mathbf{V}^{-1}_k(\mathbf{y}_k-\mathbf{h}(\mathbf{x}_k))\nonumber\\
&-\mfrac{1}{2} (\mathbf{x}_k-\mathbf{m}^{-}_k)^{\text{T}}({\mathbf{P}^{-}_{k}})^{-1}(\mathbf{x}_k-\mathbf{m}^{-}_k)\big\}\label{eqn_vb_5}
\end{align} 
where
\begin{equation}
\mathbf{V}^{-1}_k={\textbf{R}^{-1}_k}\left(diag\big(\big\langle\bm{\mathcal{I}}_k\big\rangle_{q(\bm{\mathcal{I}}_k)}\big)\right)
\label{eqn_vb_6}
\end{equation}
Using the results of Gaussian (Kalman) filter, $q(\mathbf{x}_k)$ can be approximated with a Gaussian distribution, $ {\mathcal{N}}\left(\mathbf{x}_k|\mathbf{m}^{+}_k,\mathbf{P}^{+}_{k}\right)$, with parameters given as
\begin{align}
\mathbf{m}^{+}_k&=\mathbf{m}^{-}_k+\mathbf{K}_k
(\mathbf{y}_k-\bm{\mu}_k)	\label{eqn_vb_7}\\
\mathbf{P}^{+}_{k}&=\mathbf{P}^{-}_{k}-\mathbf{C}_k\mathbf{K}^\text{T}_k\label{eqn_vb_8}
\end{align}
where {$\mathbf{m}^{-}_k$ and $\mathbf{P}^{-}_k$ denote the predicted mean and covariance matrix of a Kalman/Gaussian filter respectively and $\mathbf{m}^{+}_k$ and $\mathbf{P}^{+}_k$ denote the updated mean and covariance matrix of a Kalman/Gaussian filter respectively at the time step $k$ with}
\begin{align}
\mathbf{K}_k&=\mathbf{C}_k (\mathbf{U}_k+\mathbf{V}_k)^{-1}\nonumber\\
&=\mathbf{C}_k (\mathbf{V}^{-1}_k{-}\mathbf{V}^{-1}_k(\mathbf{I}+\mathbf{U}_k\mathbf{V}^{-1}_k)^{-1}\mathbf{U}_k\mathbf{V}^{-1}_k)\nonumber\\
\bm{\mu}_k&=\int \mathbf{h}(\mathbf{x}_k)\  {\mathcal{N}}\left(\mathbf{x}_k|\mathbf{m}^{-}_{k},\mathbf{P}^{-}_{k}\right)	d\mathbf{x}_k\nonumber\\
\mathbf{U}_k&=\int(\mathbf{h}(\mathbf{x}_k)-\bm{\mu}_k)(\mathbf{h}(\mathbf{x}_k)-\bm{\mu}_k)^{\text{T}}{\mathcal{N}}({\mathbf{x}_k}|\mathbf{m}^{-}_k,\mathbf{P}^{-}_{k})d\mathbf{x}_k\nonumber\\
\mathbf{C}_k&=\int(\mathbf{x}_k-\mathbf{m}^{-}_{k})(\mathbf{h}(\mathbf{x}_k)-\bm{\mu}_k)^{\text{T}}{\mathcal{N}}({\mathbf{x}_k}|\mathbf{m}^{-}_{k},\mathbf{P}^{-}_{k})d\mathbf{x}_k\nonumber
\end{align}
Note that the state estimates are updated with a modified measurement noise covariance $\mathbf{V}_k$ depending on the detection of outliers. When $\big\langle\bm{\mathcal{I}}_k\big\rangle_{q(\bm{\mathcal{I}}_k)}=\bm{1}$, corresponding to the ideal detection of no outlier, \eqref{eqn_vb_7}-\eqref{eqn_vb_8} become the standard Gaussian filtering equations. Likewise, when any $i$th entry of $\big\langle\bm{\mathcal{I}}_k\big\rangle_{q(\bm{\mathcal{I}}_k)}$ gets close to zero, the $i$th column of $\mathbf{K}_k$ approaches zero leading to the rejection of the corresponding $i$th measurement during inference. 

\subsection{Derivation of $q(\bm{\mathcal{I}}_k)$}
Using \eqref{eqn_vb_4} we can write
\begin{align}
q(\bm{\mathcal{I}}_k) & \propto \prod_{i=1}^{m} \frac{\sqrt{{\mathcal{I}}^i_k}}{\sqrt{2 \pi {\mathrm{R}^{ii}_k}}} \mathrm{exp}\left( {-}\frac{\mathrm{W}^{ii}_k {{\mathcal{I}}}^i_k }{2\mathrm{R}^{ii}_k}\right)  \nonumber \\  &\big((1-{\theta^i_k}) \delta({{{\mathcal{I}}}^i_k}-\epsilon)+{\theta^i_k}\delta( {{{\mathcal{I}}}^i_k}-1)\big)\label{eqn_vb_9}
\end{align} 
where $\mathrm{W}^{ii}_k$ is given as
\begin{equation}
\mathrm{W}^{ii}_k=\big\langle(\mathrm{y}^i_k-\mathrm{h}^i(\mathbf{x}_k))^2\big\rangle_{q({\mathbf{x}}_k)} 
\label{eqn_vb_9b}
\end{equation} 
We can further write 
\begin{align}
q(\bm{\mathcal{I}}_k)&=\prod_{i=1}^{m}q({{\mathcal{I}}}^i_k)=\prod_{i=1}^{m} (1-{\Omega^i_k}) \delta({{{\mathcal{I}}}^i_k}-\epsilon)+{\Omega^i_k}\delta( {{{\mathcal{I}}}^i_k}-1)
\label{eqn_vb_11}
\end{align} 
It immediately follows that $q({{\mathcal{I}}}^i_k)$ is a Bernoulli distribution with the following probabilities. 
\begin{align}
\Omega^i_k &= c  \ \theta^i_k   \mathrm{exp}(-\frac{\mathrm{W}^{ii}_k}{2\mathrm{R}^{ii}_k}) \label{t1} \\
1-\Omega^i_k &= c \ (1-\theta^i_k)\sqrt{\epsilon} \mathrm{exp}(-\frac{\mathrm{W}^{ii}_k \epsilon}{2\mathrm{R}^{ii}_k})   \label{t2}
\end{align} 
%
%
where $c$ is a proportionality constant. Using \eqref{t1} and \eqref{t2}, we obtain 

\begin{align}
\Omega^i_k &=\frac{1}{1+{\sqrt{\epsilon}}(\frac{1}{\theta^i_k}-1){\mathrm{exp}(\frac{\mathrm{W}^{ii}_k}{2\mathrm{R}^{ii}_k} (1-\epsilon))}}\label{eqn_vb_12}  
\end{align}

\subsection{Choice of the parameters $\theta^i_k$ and $\epsilon$}
For successful VB inference the choice of the parameters $\theta^i_k$ and $\epsilon$ is important. Note that in \eqref{eqn_vb_12} we cannot set $\theta^i_k$ equal to $0$ or $1$ and $\epsilon$ as exactly 0. Otherwise, the parameter $\Omega^i_k$ becomes independent of $\mathbf{W}_k$, i.e. the moment of $q(\mathbf{x}_k)$, making the VB updates impossible. We propose to set a neutral value of 0.5 or an uninformative prior for $\theta^i_k$. Moreover, a value close to zero is proposed for $\epsilon$ since exact value of 0 denies the VB updates. Note that the use of uninformative prior has also been proposed in the literature for designing outlier-robust filters assuming no prior information about the existence of outliers \cite{8869835}.

\subsection{Predictive distribution $p(\mathbf{x}_{k}|\mathbf{y}_{1:{k-1}})$}
Lastly, to complete the recursive inference process, we need to obtain the predictive distribution $p(\mathbf{x}_{k}|\mathbf{y}_{1:{k-1}})$ from the posterior distribution at the previous instant $p(\mathbf{x}_{k-1}|\mathbf{y}_{1:{k-1}})$. With the posterior distribution $p(\mathbf{x}_{k-1}|\mathbf{y}_{1:{k-1}})$ approximated as Gaussian $q(\mathbf{x}_{k-1})\approx {\mathcal{N}}\left(\mathbf{x}_{k-1}|\mathbf{m}^{+}_{k-1},\mathbf{P}^{+}_{k-1}\right)$, we use the Gaussian filtering results to write $p(\mathbf{x}_k|\mathbf{y}_{1:{k-1}})\approx {\mathcal{N}}\left(\mathbf{x}_k|\mathbf{m}^{-}_k,\mathbf{P}^{-}_{k}\right)$ where
\begin{align}
\mathbf{m}^{-}_{k}&=\int \mathbf{f}(\mathbf{x}_{k-1})\  {\mathcal{N}}\left(\mathbf{x}_{k-1}|\mathbf{m}^{+}_{k-1},\mathbf{P}^{+}_{k-1}\right)	d\mathbf{x}_{k-1}\label{eqn_vb_13}\\
\mathbf{P}^{-}_{k}&=\int\Big((\mathbf{f}(\mathbf{x}_{k-1})-\mathbf{m}^{-}_{k})(\mathbf{f}(\mathbf{x}_{k-1})-\mathbf{m}^{-}_{k})^{\text{T}}\nonumber\\&~~~~\times {\mathcal{N}}({\mathbf{x}_{k-1}}|\mathbf{m}^{+}_{k-1},\mathbf{P}^{+}_{k-1})\Big)d\mathbf{x}_{k-1}+\mathbf{Q}_{k-1}\label{eqn_vb_14}
\end{align}
The resulting selective observations-rejecting (SOR) filter is presented in Algorithm \ref{Algo1}.

\begin{algorithm}[ht!]
\SetAlgoLined
Initialize\ $\mathbf{m}^{+}_0,\mathbf{P}^{+}_0,\mathbf{Q}_k,\mathbf{R}_k$;

\For{$k=1,2...K$}{
Evaluate $\mathbf{m}^{-}_k,\mathbf{P}^{-}_k$ with \eqref{eqn_vb_13} and \eqref{eqn_vb_14}\;
Initialize $\theta^i_k$, the convergence threshold $\tau$, $\delta=\tau+1,\text{the iteration index} \ l=1 ,{\mathbf{V}^{-1}_k}^{(0)}=\mathbf{R}^{-1}_k$\;
Evaluate ${\mathbf{m}^{+}_k}^{(0)}$ and  ${\mathbf{P}^{+}_k}^{(0)}$ with \eqref{eqn_vb_7} and \eqref{eqn_vb_8}\;
\While{$\delta>\tau$}{
Update ${\mathrm{W}^{ii}_k}^{(l)}$ with \eqref{eqn_vb_9b} and ${\Omega^i_k}^{(l)}$ with \eqref{eqn_vb_12} for each $i$\;
Update ${\mathbf{V}^{-1}_k}^{(l)}$ with  \eqref{eqn_vb_6} and ${\mathbf{m}^{+}_k}^{(l)}$ and  ${\mathbf{P}^{+}_k}^{(l)}$ with \eqref{eqn_vb_7} and \eqref{eqn_vb_8}\;
Evaluate $\delta={\|{\mathbf{m}^{+}_k}^{(l)}-{\mathbf{m}^{+}_k}^{(l-1)}\|}/{\|{\mathbf{m}^{+}_k}^{(l-1)}\|}$\;
$l=l+1$\;
}
${\mathbf{m}^{+}_k}={\mathbf{m}^{+}_k}^{(l-1)}$ and ${\mathbf{P}^{+}_k}={\mathbf{P}^{+}_k}^{(l-1)}$\;
}
\caption{The proposed SOR filter}
\label{Algo1}
\end{algorithm}

\section{Performance Evaluation}\label{Eval}
\subsection{VB-based Outlier-robust Filters Under Consideration}
\subsubsection*{Standard methods}
We resort to the unscented Kalman filter (UKF) as the basic inferential engine, approximating the Gaussian integrals using the unscented transform \cite{wan2001unscented}, for all the considered methods. Therefore, we name the proposed method as selective observations-rejecting UKF (SOR-UKF). Similarly, the other VB-based outlier-robust baseline filters are called as recursive outlier-robust UKF (ROR-UKF) \cite{6349794}, switching error model UKF (SEM-UKF) \cite{8869835} and outlier-detecting UKF (OD-UKF) \cite{8398426}. The choice of using UKF allows us to propose a computationally efficient modification of our devised filter based on an available efficient implementation of UKF for handling high-dimensional uncorrelated measurements reported in literature \cite{mcmanus2012serial}.

\subsubsection*{Proposed modifications}
Since the baseline methods, in their original form, do not treat the outliers selectively for each dimension (as these are under-parameterized in terms of outlier detection for each dimension), we modify each of these for such behavior. In particular, the measurements from each sensor at each time step are proposed to be used in turn i.e. the final estimate using any sensor's observation serves as the prediction for the update using the next sensor's measurement. Accordingly, the modified baseline algorithms are referred to as mROR-UKF, mSEM-UKF, and mOD-UKF.

Moreover, we propose using an alternate implementation of the proposed SOR-UKF, referred to as mSOR-UKF with lower complexity. Since $\mathbf{V}^{-1}_k$ is diagonal, the basic inferential UKF used within the proposed SOR-UKF can be modified from the standard parallel sigma point Kalman filter (P-SPKF) to the serial sigma point Kalman filter (S-SPKF) \cite{mcmanus2012serial} for use in mSOR-UKF. Note that S-SPKF can be also used in conjunction with ROR-UKF, SEM-UKF and OD-UKF, as originally reported, but these do not treat outliers selectively for each dimension hence compromising the estimation quality. mROR-UKF, mSEM-UKF and mOD-UKF, on the other hand, have analogous structures to the serial re-draw sigma point Kalman filter (SRD-SPKF) in \cite{mcmanus2012serial}.
\subsection{Theoretical Computational Complexity}
\begin{table}[h!]
\caption{Theoretical complexity of different learning-based methods treating outliers selectively}
\centering
\begin{tabular}{@{}lll@{}}
\toprule
\ \ \textbf{{Method}}  & \textbf{{Complexity}} & \textbf{Computational Analogue}\\ \midrule
\ \ {SOR-UKF}           & {$\mathcal{O}{(}(m+n)^3{)}$} & \quad \quad \quad P-SPKF \\
\ \ {mOD/mSEM/mROR-UKF} & {$\mathcal{O}(m n^3)$}  & \quad \quad \quad SRD-SPKF  \\
\ \ {mSOR-UKF}          &  {$\mathcal{O}(n^2(m+n))$} & \quad \quad \quad S-SPKF \\ \bottomrule
\end{tabular}
\label{Tab1}
\end{table}

The proposed SOR-UKF in the original form involves matrix inversions for Kalman gain evaluation and sigma points are drawn after processing the entire vector of measurements, resultingly it has a theoretical complexity of $\mathcal{O}{(}(m+n)^3{)}$. In mROR-UKF, mSEM-UKF, and mOD-UKF after updating the state estimate using one sensor's measurement, sigma points from the updated state distribution are regenerated, in VB iterations until convergence, for processing the next sensor's measurement during the same sampling interval. During updates, only scalar inversions are required for Kalman gain computations. Consequently, these methods have a theoretical complexity of $\mathcal{O}(m n^3)$. In mSOR-UKF, S-SPKF is used which has complexity of $\mathcal{O}(n^2(m+n))$. Moreover, all the $\mathrm{W}^{ii}_k$ terms can be computed with complexity not more than $m n$, we can implement mSOR-UKF with an overall theoretical complexity of $\mathcal{O}(n^2(m+n))$. Table~\ref{Tab1} summarizes the theoretical complexity of different learning-based methods treating multiple outliers selectively along with their standard analogous counterparts. Note that since the theoretical complexity depends on the exact functional forms of $\textbf{f}(.)$ and $\textbf{h}(.)$, we assume in the analysis that each entry of the functional mapping requires constant time independent of $n$. The assumption is true for several practical cases e.g. range data of a moving target from sensors depends on the position of target only.

\subsection{Simulation Results}
\subsubsection*{Simulation setup}
In all the simulations and experimental scenarios, we use Matlab on an Intel i7-8550U processor powered computer and consider SI units. First, we choose to simulate a target tracking application for performance evaluation. 

\begin{figure}[!h]
\centering
\includegraphics[width=0.9\columnwidth]{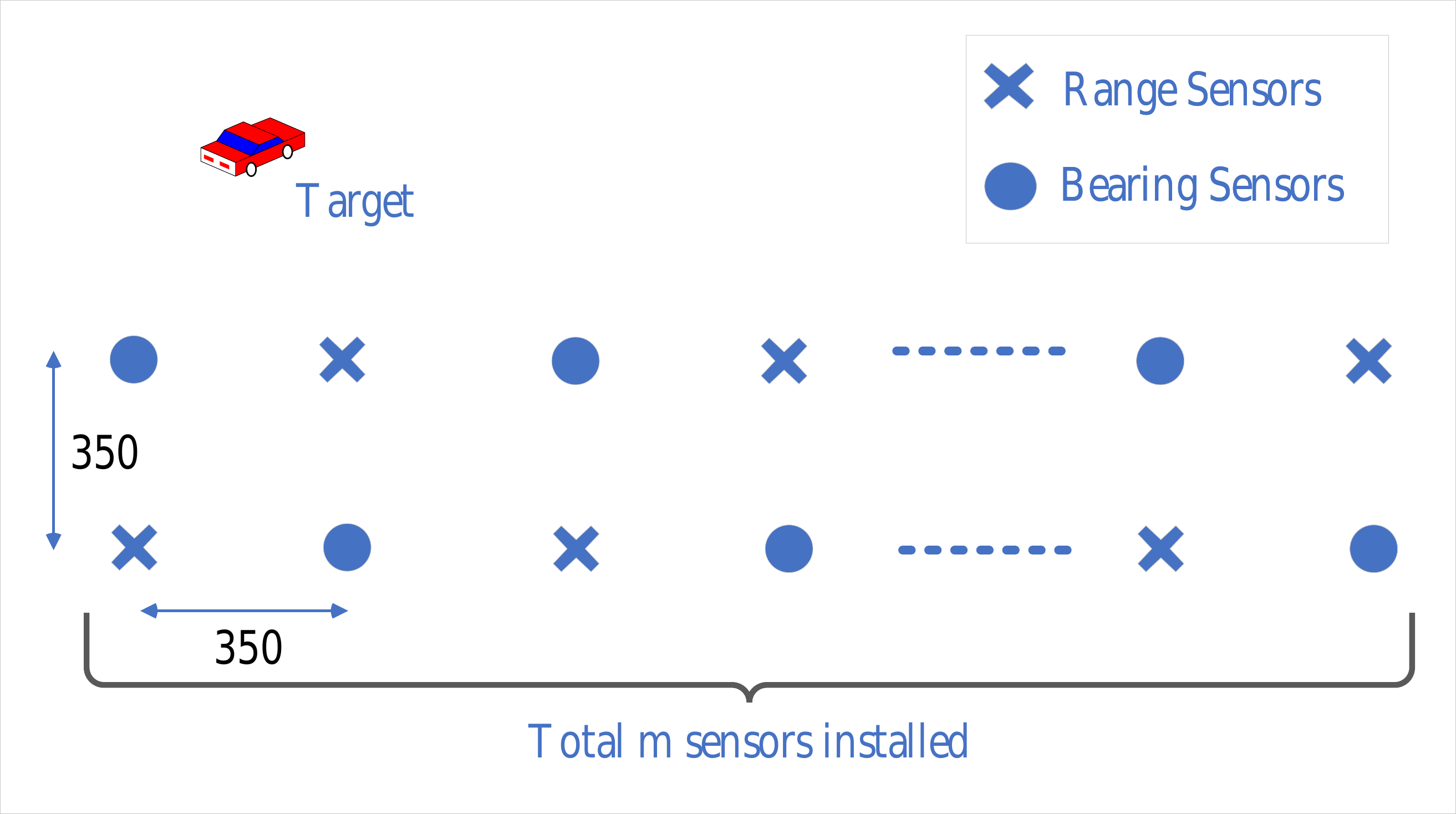}
\caption{Setup of the target tracking test example}
\label{results_setup}
\end{figure}

Fig.~\ref{results_setup} shows the testing setup for a target tracking example we consider in the simulations for performance evaluation. Maneuvering targets with unknown turn rates are commonly modeled as 
\begin{align}
\mathbf{x}_k &= \begin{pmatrix} \text{1} & \frac{\text{sin}(\omega_{k}\Delta t)}{\omega_{k}} & \text{0} &  \frac{\text{cos}(\omega_{k}\Delta t)-\text{1}}{\omega_{k}} & \text{0} \\  \text{0} & \text{cos}(\omega_{k}\Delta t) &  \text{0} & -\text{sin}(\omega_{k}\Delta t) & \text{0}\\  \text{0} & \frac{\text{1}-\text{cos}(\omega_{k}\Delta t)}{\omega_{k}} & \text{1} &\frac{\text{sin}(\omega_{k}\Delta t)}{\omega_{k}} & \text{0} \\  \text{0} &  \text{\text{sin}}(\omega_{k}\Delta t) &  \text{0} &  \text{cos}(\omega_{k}\Delta t) & \text{0}\\  \text{0} & \text{0} & \text{0} & \text{0} &\text{1} \end{pmatrix} \mathbf{x}_{k-\text{1}} + \mathbf{q}_{k-\text{1}} \label{eqn_res1} 
\end{align}
where the state vector $\mathbf{x}_k= [a_k,\dot{{a_k}},b_k,\dot{{b_k}},\omega_{k}]^\text{T}$
contains the 2D position coordinates \(({a_k} , {b_k} )\), the corresponding velocities \((\dot{{a_k}} , \dot{{b_k}} )\), the angular velocity $\omega_{k}$ of the target at time instant $k$, \( \Delta t \) is the sampling period, and $\mathbf{q}_{k-\text{1}} \sim N\left(0,\mathbf{Q}_{k-\text{1}}.\right)$ with $\mathbf{Q}_{k-\text{1}}$ given, in terms of scaling parameters $\eta_1$ and $\eta_2$, as \cite{8398426}
\begin{equation}
\mathbf{Q}_{k-\text{1}}=\begin{pmatrix} \eta_1 \mathbf{M} & 0 & 0\\0 &\eta_1 \mathbf{M}&0\\0&0&\eta_2
\end{pmatrix}, \mathbf{M}=\begin{pmatrix} {\triangle t}^3/3 & {\triangle t}^2/2\\{\triangle t}^2/2 &{\triangle t}
\end{pmatrix}\nonumber
\end{equation}

We consider that angle and range readings are obtained from sensors, installed around a rectangular area, at $m$ different locations. A total of $m/2$ independent sensors are used to provide angle readings where its $j$th sensor is present at the 2D coordinate $(a^{\theta_j}=350(j-1),b^{\theta_j}=350(j\mod 2))$. Similarly, range measurements are obtained from the other $m/2$ independent sensors where its $j$th sensor is located at $(a^{\rho_j}=350(j-1),b^{\rho_j}=350\ ((j-1)\mod2))$. 



\begin{figure}[h!]
\centering
\hspace*{.2cm}\includegraphics[width=.95\linewidth]{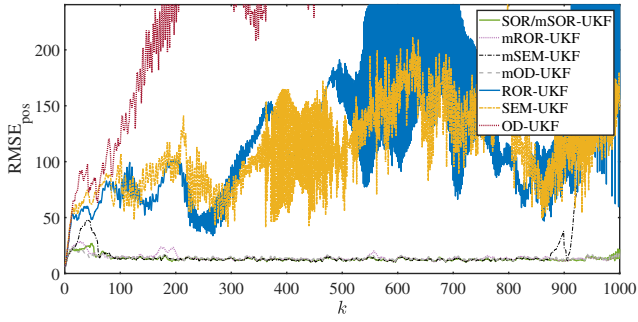}
\caption{$\text{RMSE}_{\text{pos}}$ vs. $k$ \\ ($\lambda=0.3, \gamma\sim \mathcal{U}(100,1000)$)}
\label{fig:lambda6mixgauss_RMSE1000}
\end{figure}

\begin{figure}[!h]
\centering
\includegraphics[width=1.05\columnwidth]{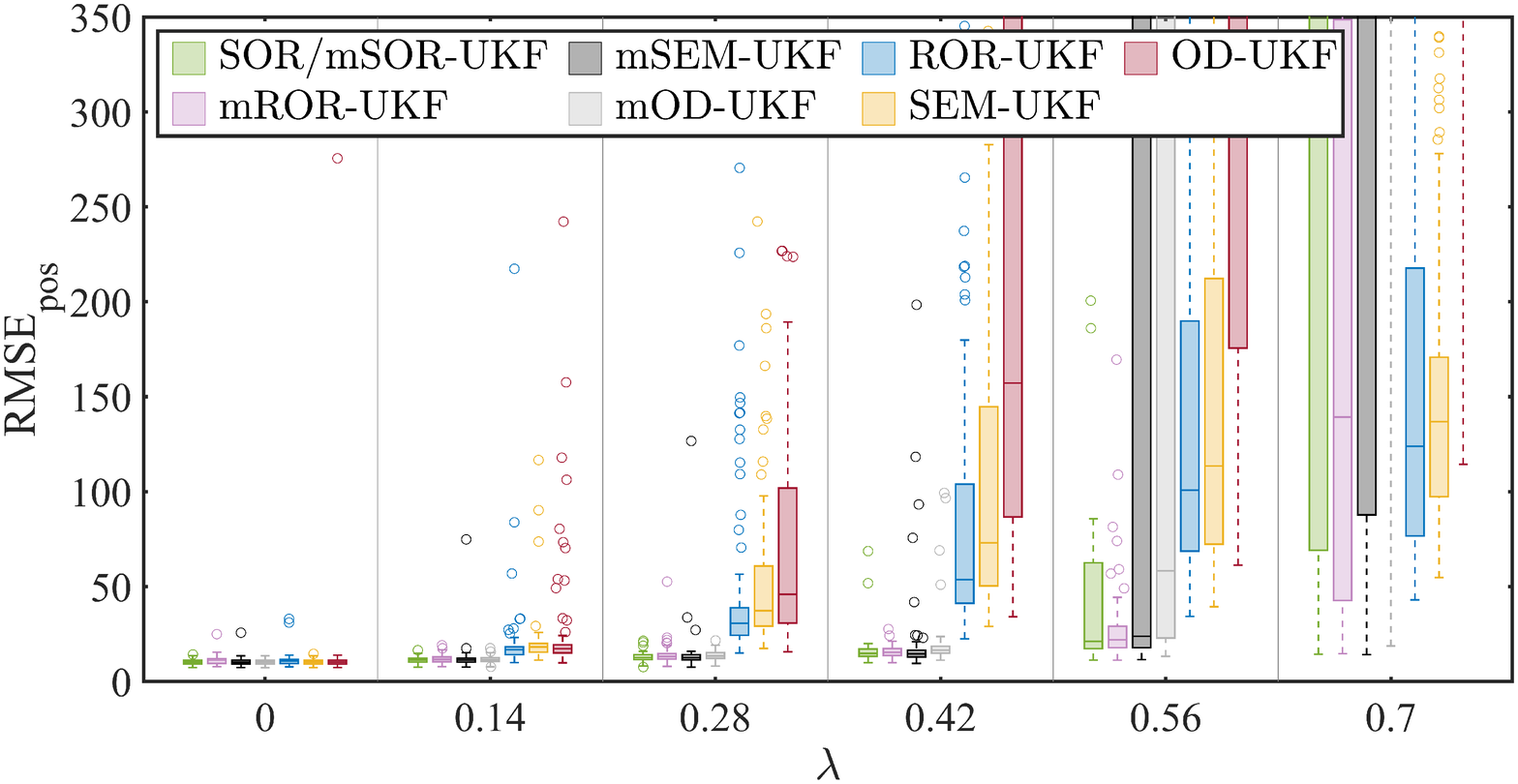}
\caption{$\text{RMSE}_{\text{pos}}$ vs. $\lambda$ \\ ($\gamma\sim \mathcal{U}(100,1000)$)}
\label{results_2}
\end{figure}

\begin{figure}[!h]
\centering
\includegraphics[width=1.05\columnwidth]{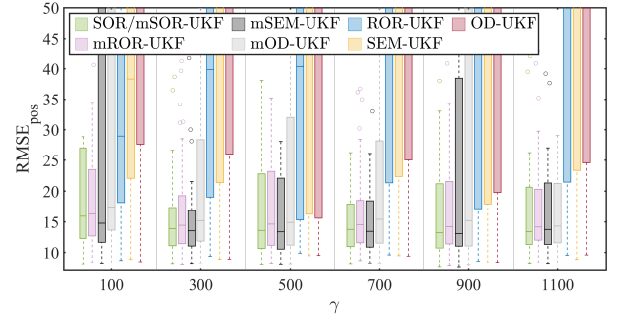}
\caption{$\text{RMSE}_{\text{pos}}$ vs. $\gamma$ \\ ($\lambda\sim \mathcal{U}(0,0.7)$)} 
\label{results_1}	
\end{figure}

\begin{figure}[!h]
\centering
\includegraphics[width=1.05\linewidth]{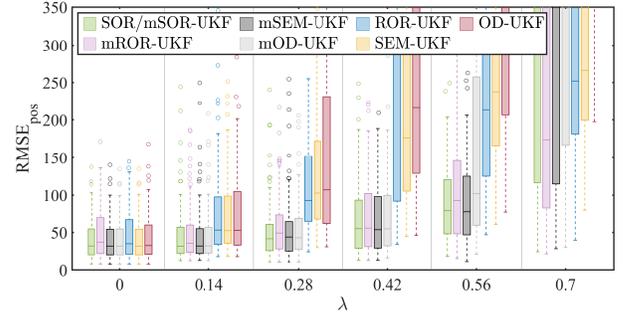}
\caption{$\text{RMSE}_{\text{pos}}$ vs. $\lambda$ ($\gamma\sim \mathcal{U}(100,1000)$, ${\sigma_\theta}\sim \mathcal{U}(1.7\times10^{-3},3.5\times10^{-2})$, ${\sigma_\rho}\sim \mathcal{U}(5,50)$)}
\label{fig:lambda6mixgauss_sigma}
\end{figure}

\begin{figure}[h!]
\centering
\includegraphics[width=1.05\linewidth]{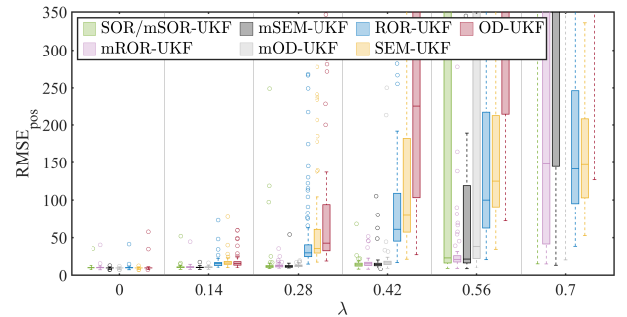}
\caption{$\text{RMSE}_{\text{pos}}$ vs. $\lambda$ ($\gamma\sim \mathcal{U}(100,1000)$, $
\epsilon \sim \mathcal{U}(10^{-7},10^{-3})$, $\theta^i_k\sim \mathcal{U}(0.05,0.95)$)}
\label{fig:lambda6mixgauss_randeps}
\end{figure}

\begin{figure}[h!]
\centering
\includegraphics[width=1.05\linewidth]{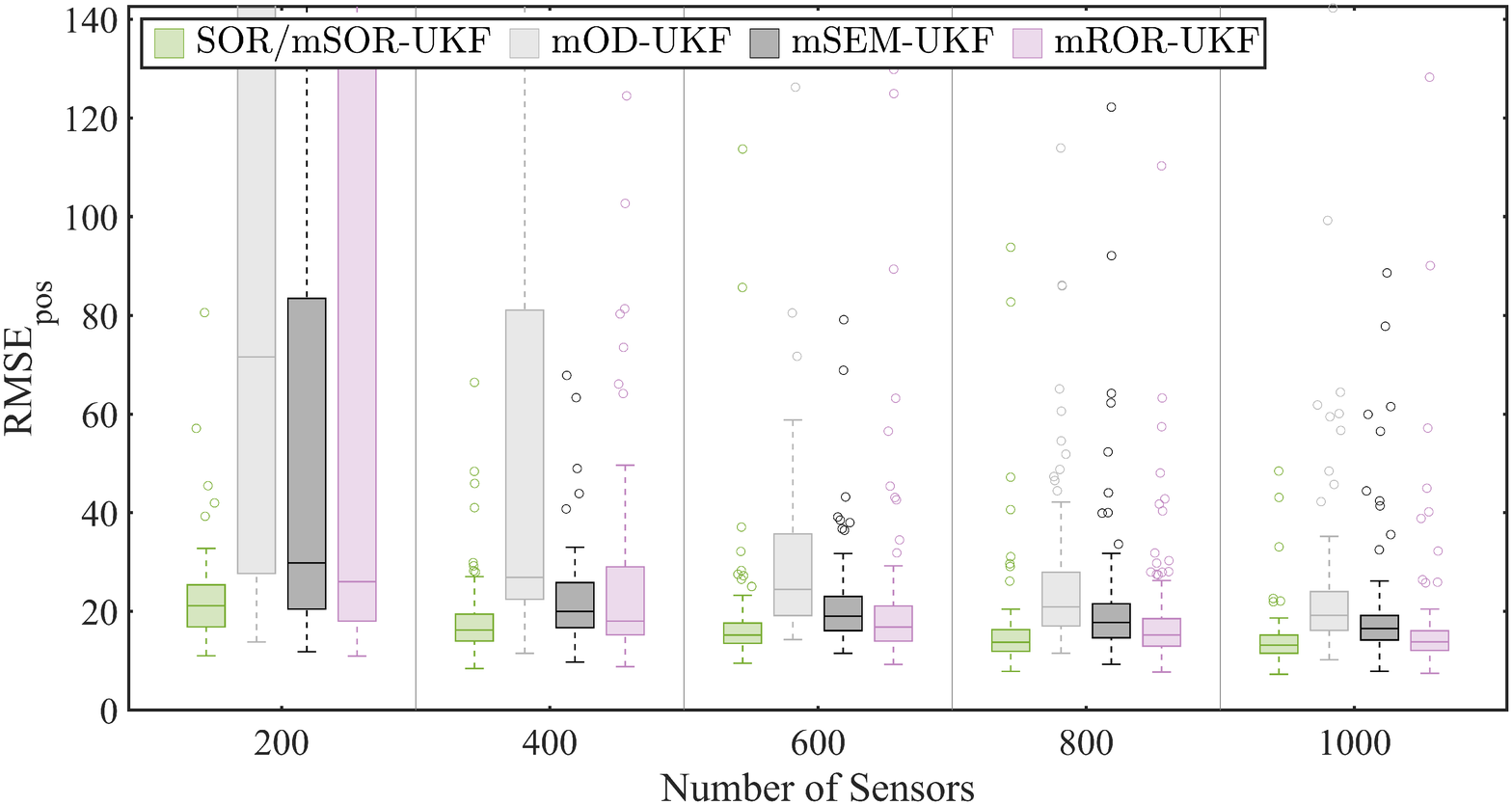}
\caption{$\text{RMSE}_{\text{pos}}$ vs. $m$ ($\lambda=0.9$, $\gamma\sim \mathcal{U}(100,1000)$)}
\label{fig:lambda6mixgauss_lam_pntnine}
\end{figure}

\begin{figure}[h!]
\centering
\includegraphics[width=1.05\linewidth]{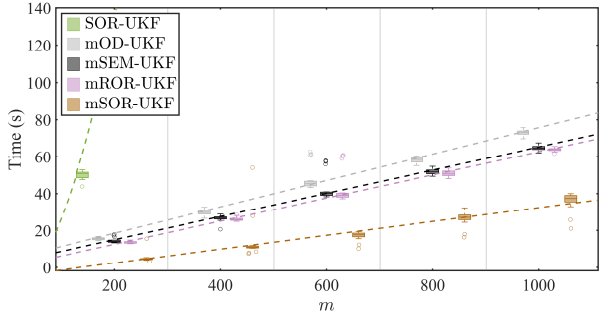}
\caption{$\text{Computational overhead}$ vs. $m$ ($\lambda=0.9$,  $\gamma\sim \mathcal{U}(100,1000)$)}
\label{fig:complexity}
\end{figure}

{For performance evaluation, two subcases of how measurements from different sensors get disturbed are considered.} {First, we suppose outliers explicitly add in the observations randomly. Subsequently, we evaluate the case of missing data. 

\subsubsection*{{Outliers}}	
We suppose a Gaussian mixture to model the additive effect of outliers commonly used for performance evaluation of robust filtering methods\cite{8869835,8398426}.} The measurement vector $\mathbf{y}_k$  considering the {additive} effect of independently occurring outliers in different dimensions {can be expressed} as
\begin{equation}
\mathbf{y}_k=\mathbf{s}_k + \mathbf{o}_k \label{eqn_res2}
\end{equation}
where $\mathbf{s}_k=[s^{\theta_1}_k...s^{\theta_{{m}/{2}}}_k,s^{\rho_1}_k...s^{\rho_{m/2}}_k] ^\text{T}$ and $\mathbf{o}_k = [o^{\theta_1}_k...o^{\theta_{m/2}}_k,o^{\rho_1}_k...o^{\rho_{m/2}}_k]^\text{T}$
denote the noise-free and noisy components of $\mathbf{y}_k$ respectively. The entries of $\mathbf{s}_k$, $s^{\theta_j}_k$ and $s^{\rho_j}_k$, which denote the noise-free values corresponding to $j$th bearing and range sensors respectively are given as

\begin{align}
s^{\theta_j}_k&=\text{atan2}(b_k-b^{\theta_j},a_k-a^{\theta_j})\label{eqn_res3}\\
s^{\rho_j}_k&=\left({(a_k-a^{\rho_j})^{\text{2}} + (b_k-b^{\rho_j})^{\text{2}}}\right)^{\tfrac{1}{2}}\label{eqn_res3b}
\end{align}
Similarly, the entries of $\mathbf{o}_k$, $o^{\theta_j}_k$ and $o^{\rho_j}_k$, control the noise content in the measurements from the $j$th bearing and range sensors. $\mathbf{o}_k$ is considered to obey the following distribution
\begin{align}
p(\mathbf{o}_k)&=\prod_{j=1}^{m/2}\big(\lambda\   {\mathcal{N}}(o^{\theta_j}_k|0,\gamma\ {\sigma_\theta}^2)+(1-\lambda)\  {\mathcal{N}}(o^{\theta_j}_k|0,{\sigma_\theta}^2)\big)\nonumber\\ \times&\left( \lambda\  {\mathcal{N}}(o^{\rho_j}_k|0,\gamma\ {\sigma_\rho}^2)+(1-\lambda)\  {\mathcal{N}}(o^{\rho_j}_k|0, {\sigma_\rho}^2) \right)\label{eqn_res4} 
\end{align}
where ${\sigma_\theta}^2$ and ${\sigma_\rho}^2$ are the variances of nominal noise in angle and range readings respectively. The parameters $\lambda$ and $\gamma$ control the frequency and variance of an outlier in each dimension respectively. 

\subsubsection*{{Base parameters}}
The following values of parameters are used ({unless stated otherwise}): the initial state $\mathbf{x}_0= [-10000,10,5000,-5,-0.0524] ^\text{T}$, $\triangle t=1$, $\eta_1=0.1$, $\eta_2=1.75\times10^{-4}$, ${\sigma_\theta}=3.5\times10^{-3}$, ${\sigma_\rho}=10$ {and $m=6$}.
The initialization parameters of filters are: {${\mathbf{m}}^+_{0} \sim \mathcal{N}(\mathbf{x}_0,\mathbf{P}^+_{0}$), $\mathbf{P}^+_{0}=100\mathbf{Q}_{k}$}, $\epsilon=10^{-6}$ and $\theta^i_k=0.5~\forall~i$. For each method, the UKF parameters \cite{wan2001unscented} are set as $\alpha=1$, $\beta=2$ and $\kappa=0$. Moreover, we use the same convergence threshold $\tau=10^{-4}$ and convergence metric $\delta$ for all the evaluated methods. Other parameters for the rival methods are assigned values as originally reported. All the simulations are repeated with a total time duration {$K=1000$} and $100$ independent Monte Carlo (MC) runs.

{
First, we assess the tracking performance over time of different filters. Fig.~\ref{fig:lambda6mixgauss_RMSE1000} shows the Root Mean Squared Error of the position estimates $\text{RMSE}_{\text{pos}}$ over time for this scenario assuming $\lambda=0.3, \gamma\sim \mathcal{U}(100,1000)$. The methods which treat outliers selectively for each dimension i.e. SOR-UKF, mSOR-UKF, mROR-UKF, mSEM-UKF, and mOD-UKF exhibit comparable tracking performance  whereas the other filters result in larger errors.

We also assess the quality of estimates using different filters with a change in $\lambda$ with $\gamma\sim \mathcal{U}(100,1000)$. Fig.~\ref{results_2} shows the box plots of the Root Mean Squared Error of the position estimates  $\text{RMSE}_{\text{pos}}$ for this scenario. The methods dealing selectively with outliers exhibit comparable performance whereas the other filters result in a sharper rise of $\text{RMSE}_{\text{pos}}$ with an increase in $\lambda$.

Subsequently, we vary $\gamma$ and observe $\text{RMSE}_{\text{pos}}$ with $\lambda\sim\mathcal{U}(0,0.7)$. Fig.~\ref{results_1} shows the distributions of the $\text{RMSE}_{\text{pos}}$ for this case using different filters. SOR-UKF, mSOR-UKF, mROR-UKF, mSEM-UKF, and mOD-UKF demonstrate comparable performance and outperform other methods. 

In addition, we evaluate how the change in nominal noise parameters ${\sigma_\theta}$ and ${\sigma_\rho}$ affects the performance of filters. Varying $\lambda$ the change in $\text{RMSE}_{\text{pos}}$ with $\gamma\sim \mathcal{U}(100,1000)$, ${\sigma_\theta}\sim \mathcal{U}(1.7\times10^{-3},3.5\times10^{-2})$, ${\sigma_\rho}\sim \mathcal{U}(5,50)$ is depicted in the box plots in Fig.~\ref{fig:lambda6mixgauss_sigma}. We observe a similar trend as in Fig.~\ref{results_2} except that the error magnitude levels increase.

In addition, we evaluate the robustness of SOR-UKF and mSOR-UKF with change in filter parameters by assuming $\epsilon \sim \mathcal{U}(10^{-7},10^{-3})$ and $\theta^i_k\sim \mathcal{U}(0.05,0.95)$ with $\gamma\sim\mathcal{U}(100,1000)$. Varying $\lambda$ the change in $\text{RMSE}_{\text{pos}}$ is depicted in the box plots in Fig.~\ref{fig:lambda6mixgauss_randeps}. We find the proposed filters quite robust to changes in filter parameters except at large values of $\lambda$ where the SOR-UKF and mSOR-UKF exhibit larger errors. 


We also simulate a case where the effect of increase in the number of sensors on the estimation quality is observed. Fig.~\ref{fig:lambda6mixgauss_lam_pntnine} depicts the $\text{RMSE}_{\text{pos}}$ versus $m$, increased from 200 to 1000, for methods dealing each dimension selectively. We choose a high rate of outlier occurrence i.e. $\lambda=0.9$ with $\gamma\sim\mathcal{U}(100,1000)$. Owing to a large dimensionality of $m$, or more sources of information, we see that most of the algorithms result in low errors especially at higher values of $m$. We find SOR-UKF and mSOR-UKF comparatively more robust in this scenario. Note that non-selective methods exhibit higher errors so we skip them in the results.

Lastly, we compare the computational cost for the selective methods exhibiting comparable lower $\text{RMSE}_{\text{pos}}$ values. We evaluate these methods in terms of the time taken to complete the Monte Carlo simulations. Setting $\lambda=0.9$ with $\gamma\sim\mathcal{U}(100,1000)$, we vary $m$ and present the distributions of completion times of simulations using different filters in Fig.~\ref{fig:complexity}. With increasing $m$, the average completion times of the simulations increase for all the filters. We observe that the empirical computational overhead verifies the theoretical complexity. SOR-UKF exhibits cubic complexity whereas the other techniques have linear complexity in terms of $m$. The differences in the processing overhead among mROR-UKF, mSEM-UKF, and mOD-UKF depends on the their modeling parameters and inferential mechanism. Importantly, we note that with an increase in $m$, the computational costs of mROR-UKF, mSEM-UKF, and mOD-UKF rise more steeply than the mSOR-UKF as predicted by the theoretical complexity (analogous to how S-SPKF is computationally faster than the SRD-SPKF) in \cite{mcmanus2012serial}).

}

\subsubsection*{{Missing data}}

{We also evaluate the performance of different filters for the case of missing data. Missing observations can} {also be viewed as special case of outliers in a sense that the nominal model in \eqref{eqn_model_1}-\eqref{eqn_model_2} is unable to describe the observations and only the modified measurement equation \eqref{eqn_model_2-m} can with ${{\mathcal{I}}}^i_k=\epsilon$ in any particular affected dimension $i$. Using the base parameters, as in the case of outliers, we simulate this case with $\lambda$ indicating the probability of missing observations in each dimension.

\begin{figure}[h!]
\centering
\includegraphics[width=0.95\linewidth]{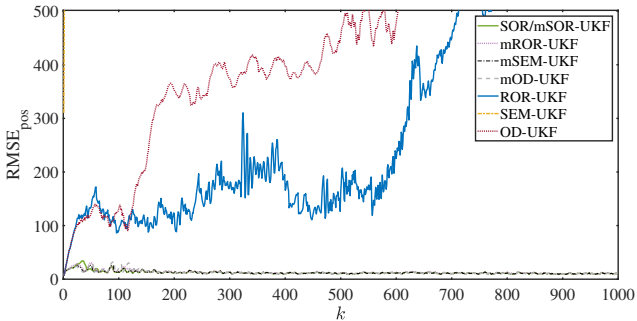}
\caption{$\text{RMSE}_{\text{pos}}$ vs. $k$ \\ $\big(\lambda=0.3$$\big)$}
\label{fig:lambda6missing_RMSE1000}
\end{figure} 

\begin{figure}[h!]
\centering
\includegraphics[width=1.05\linewidth]{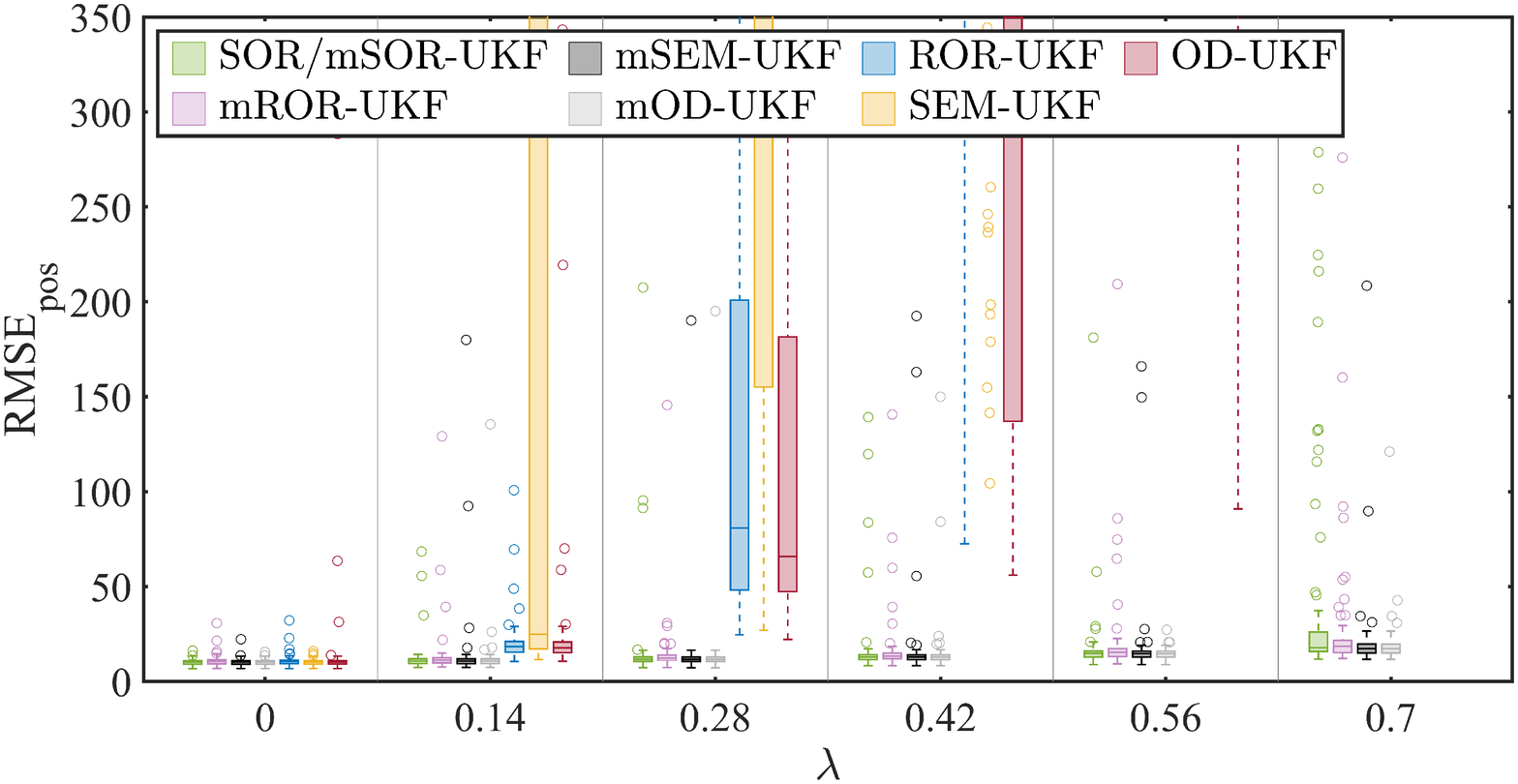}
\caption{$\text{RMSE}_{\text{pos}}$ vs. $\lambda$}
\label{fig:lambda6missing}
\end{figure}
\begin{figure}[h!]
\centering
\includegraphics[width=1.05\linewidth]{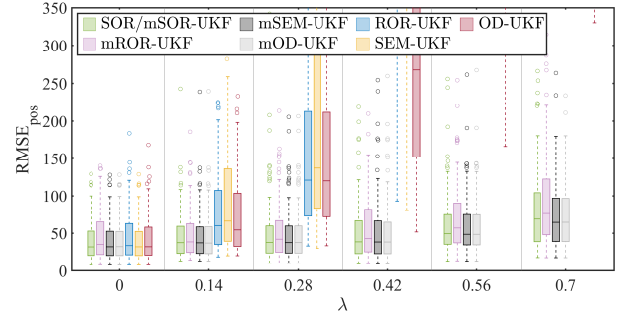}
\caption{$\text{RMSE}_{\text{pos}}$ vs. $\lambda$, ${\sigma_\theta}\sim \mathcal{U}(1.7\times10^{-3},3.5\times10^{-2})$, ${\sigma_\rho}\sim \mathcal{U}(5,50)$)}
\label{fig:lambda6missingsig}
\end{figure}

\begin{figure}[h!]
\centering
\includegraphics[width=1.05\linewidth]{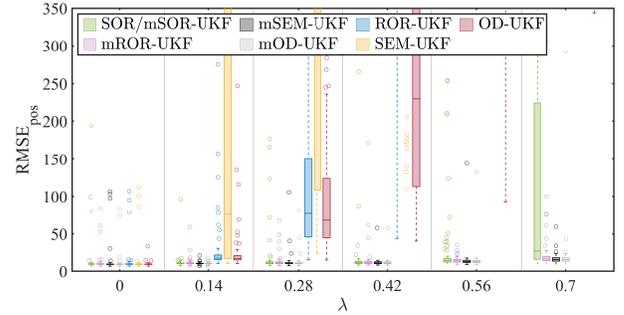}
\caption{$\text{RMSE}_{\text{pos}}$ vs. $\lambda$, $
\epsilon \sim \mathcal{U}(10^{-7},10^{-3})$, $\theta^i_k\sim \mathcal{U}(0.05,0.95)$}
\label{fig:lambda6missingrandeps}
\end{figure}

\begin{figure}[h!]
\centering
\includegraphics[width=1.05\linewidth]{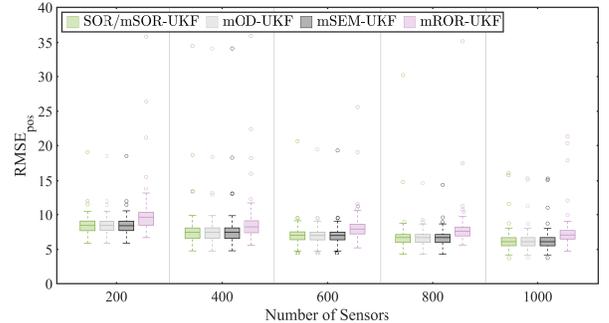}
\caption{$\text{RMSE}_{\text{pos}}$ vs. $m$ \\ ($\lambda=0.9$)}
\label{fig:missing_lam_pntnine}
\end{figure}

We observe similar results as for the case of outliers. Fig.~\ref{fig:lambda6missing_RMSE1000} shows the tracking error of different algorithms over time with $\lambda=0.3$. We again find the selective methods exhibiting lower} {errors.

\begin{figure*}[b!]
	\centering
	\begin{minipage}[b!]{0.32\textwidth}
		\centering
		\includegraphics[width=0.99\textwidth]{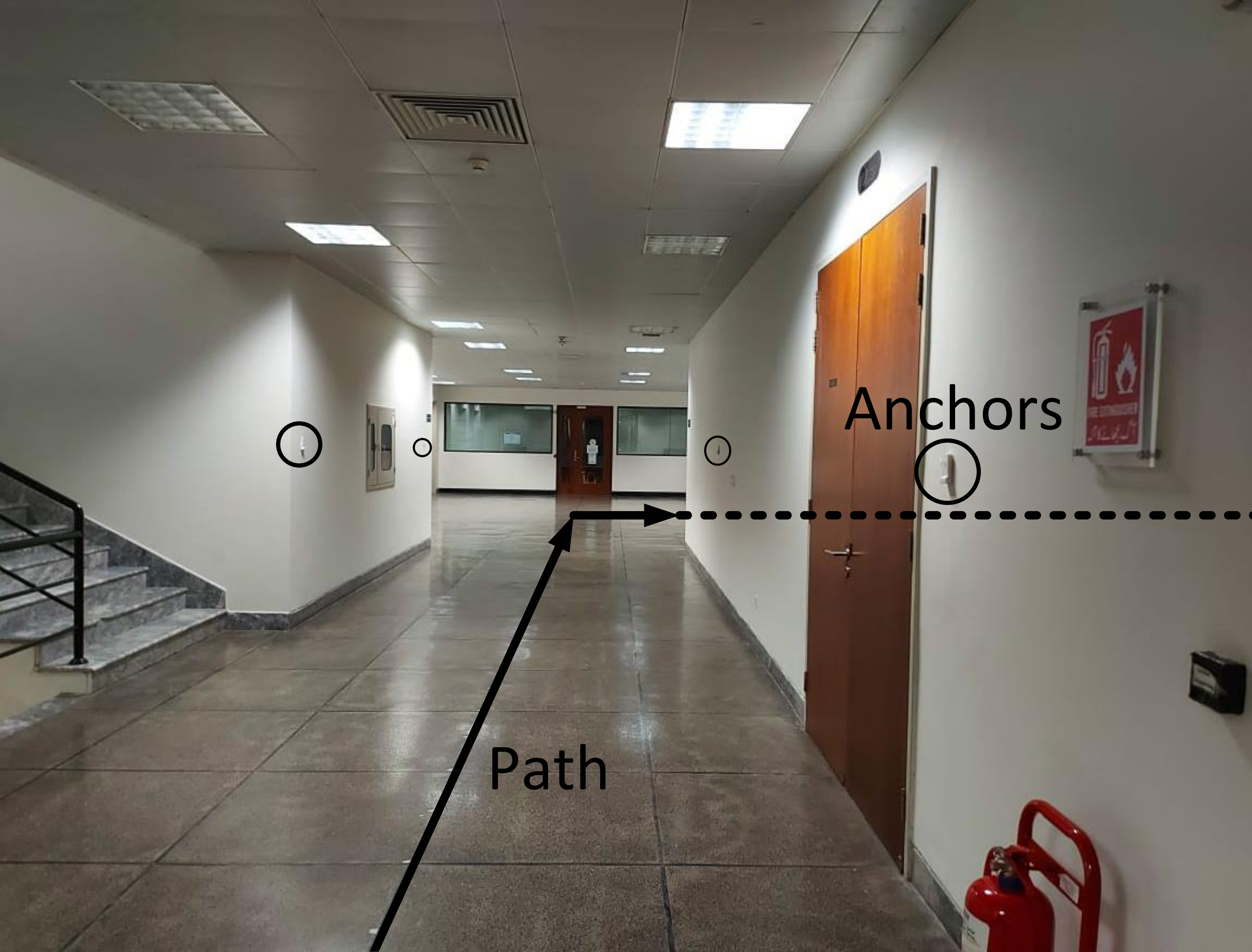} \caption{Experimentation site 1: SSE corridor}
		\label{Exp_sc_results_1}	
	\end{minipage}
	\begin{minipage}[b!]{0.32\textwidth}
		\centering
		\includegraphics[width=0.99\textwidth]{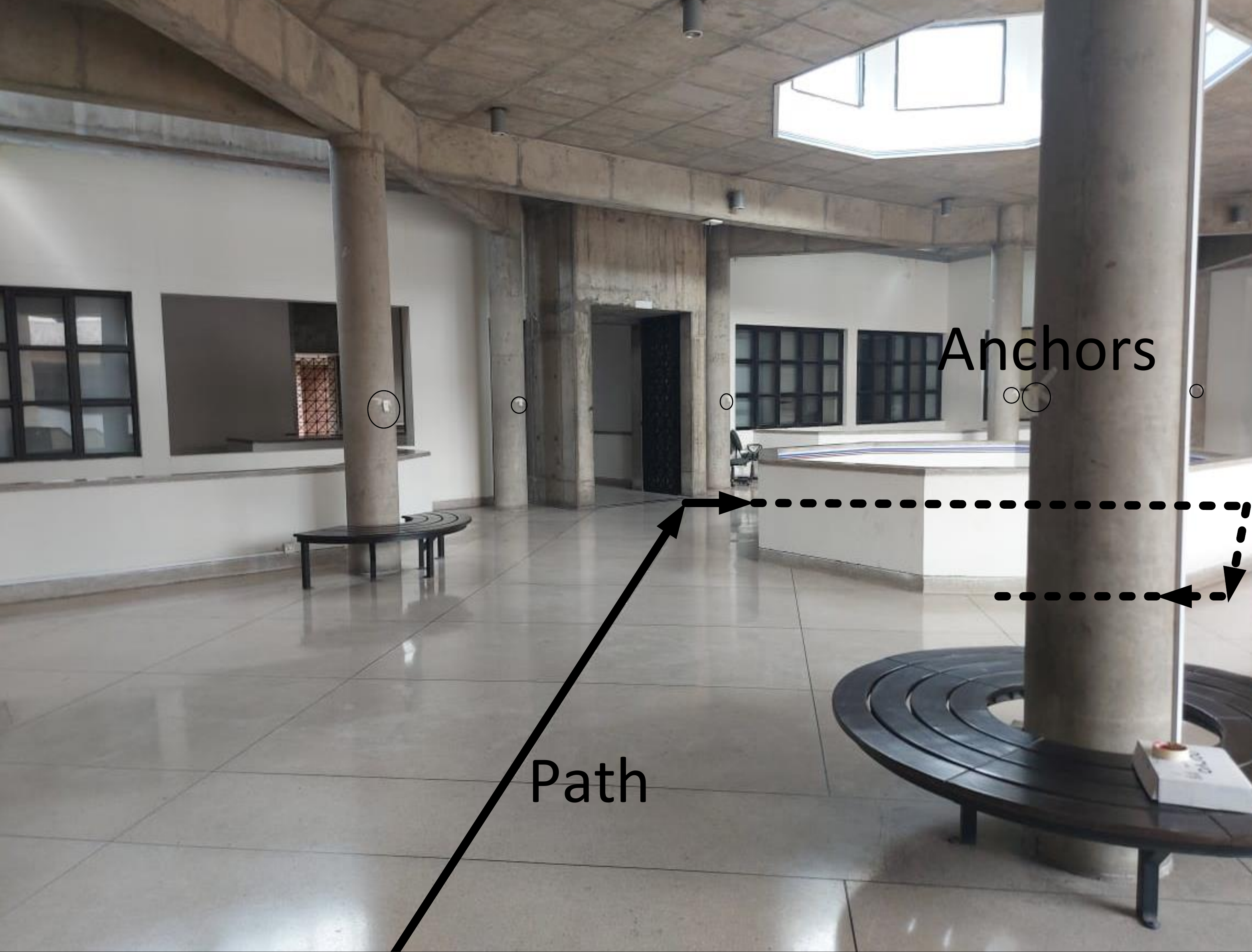}
		\caption{Experimentation site 2: AB corridor} 
		\label{Exp_sc_results_2}	
	\end{minipage}
	\begin{minipage}[b!]{0.32\textwidth}
		\centering
		\includegraphics[width=0.99\textwidth]{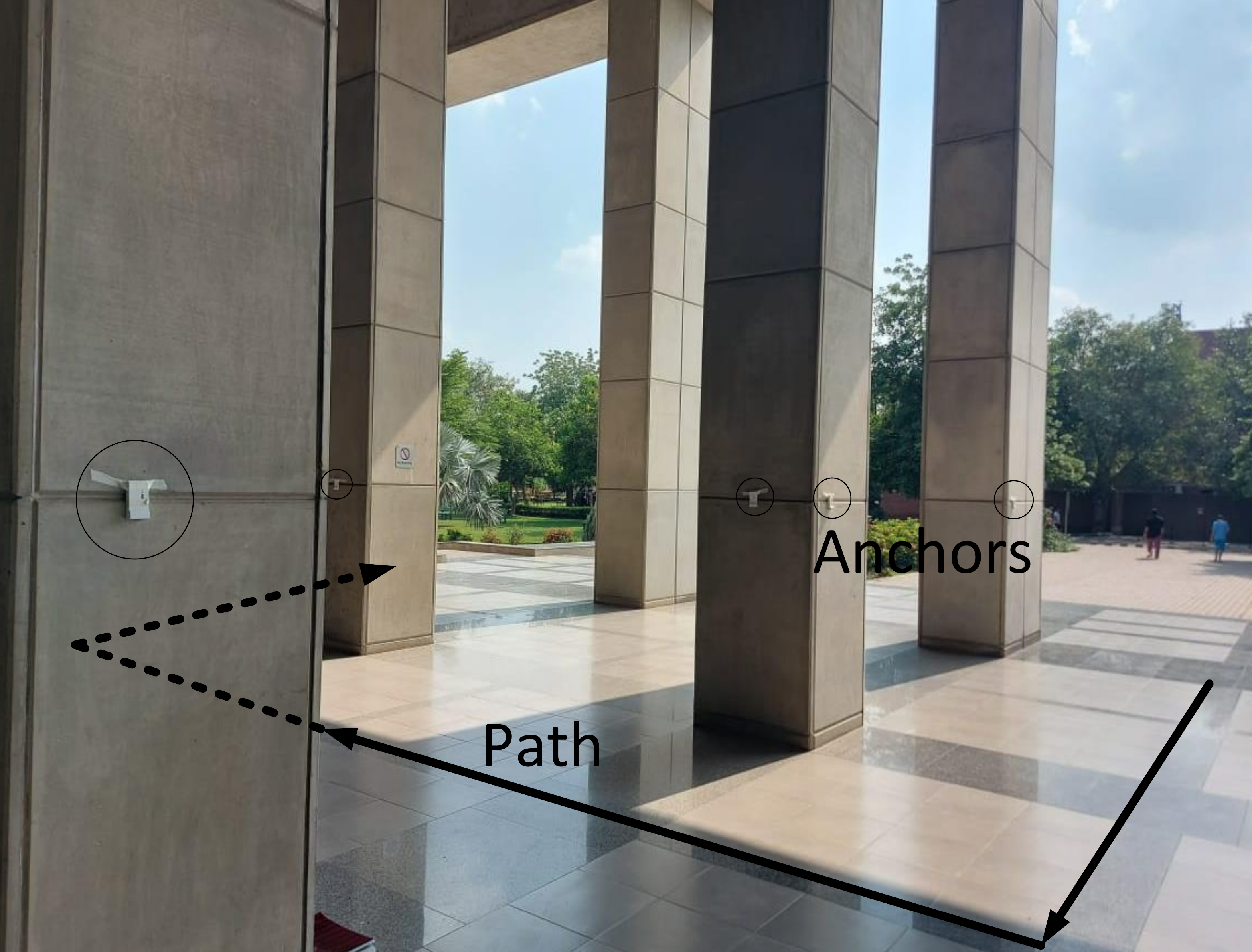}
		\caption{Experimentation site 3: SSE entrance}
		\label{Exp_sc_results_3}
	\end{minipage}
\end{figure*}

Fig.~\ref{fig:lambda6missing} shows the box plots of $\text{RMSE}_{\text{pos}}$ with change in $\lambda$ which reaffirms the superiority of methods dealing each dimension independently. However, we observe more} {degraded performances of non-selective methods compared to the case of outliers, drawn from a Gaussian mixture model, with increase in $\lambda$ especially the compensation-based methods i.e. SEM-UKF and ROR-UKF. These methods try to extract information from the measurement vector by learning the} {scaling parameters of their respective Gaussian covariance matrices given the abnormalities in data. This may be more suitable for certain applications and is not useful generally for unknown clutter or "nonsense outliers" in words of the} {authors of SEM-UKF. For more detailed discussion where such non-selective compensating models are more useful the readers are referred to the Discussion Section of SEM-UKF pg. 8 \cite{8869835}.

Fig.~\ref{fig:lambda6missingsig} shows the box plots of $\text{RMSE}_{\text{pos}}$ with change in $\lambda$ by assuming ${\sigma_\theta}\sim \mathcal{U}(1.7\times10^{-3},3.5\times10^{-2})$ and ${\sigma_\rho}\sim \mathcal{U}(5,50)$ to depict how the change in nominal noise statistics affect the estimation performance. The results have a similar trend as in Fig.~\ref{fig:lambda6missing} with a change that the error magnitude levels increase.

To test the robustness of the proposed filters with variations in filter parameters we assume $\epsilon \sim \mathcal{U}(10^{-7},10^{-3})$ and $\theta^i_k\sim \mathcal{U}(0.05,0.95)$. With variation in $\lambda$, box plots of $\text{RMSE}_{\text{pos}}$ are given in Fig.~\ref{fig:lambda6missingrandeps}. Similar to the case of outliers, we observe the proposed algorithms to be quite robust to changes in filter parameters except at larger values of $\lambda$ where these produce more estimation errors.   

As for the case of outliers in the previous case, we evaluate the comparative performance of methods dealing each dimension selectively with increase in $m$ and observe a similar trend for $\lambda=0.9$ as depicted in Fig.~\ref{fig:missing_lam_pntnine}. in this case the estimation quality remains similar for each method.

For the computational complexity we have observed a similar trend for the case of missing observations as reported for the case of outliers as reported in Fig.~\ref{fig:complexity}.

Lastly, note that a case might arise where any actual measurement is close to zero e.g. the target may be very close to any range sensor in the simulation example. For this case, the nominal model would be able to explain the data as a result the measurement would be used for inference instead of being discarded. We have also simulated this case by initializing the target very close to a range sensor's coordinates and repeating the missing observations simulations to test the robustness of the filters. For this case, we arrive at the similar results as reported above for the case of missing data simulations.}

\subsection{Experimental Results}

\subsubsection*{Experimental campaigns} 

\begin{table*}[h!]
\centering
\caption{Performance results of different VB-based outlier-robust filters for three experimental settings of indoor localization}
\begin{tabular}{lcccccc} 
\toprule
\multirow{2}{*}{} & \multicolumn{2}{c}{~Scenario 1} & \multicolumn{2}{c}{~Scenario 2} & \multicolumn{2}{c}{~Scenario 3}  \\ 
\cmidrule{2-7}
& ~RMSE~ & Mean Run Time  & ~RMSE~ & Mean Run Time & ~RMSE~ & Mean Run Time   \\ 
\midrule
mROR-UKF          &      {0.16}      &     {0.24}              &    {0.16}      &   {0.21}                 &   {9.26}       &        {0.25}             \\
mOD-UKF           &    {0.18}      &    {0.18}              &   {0.21}        &      {0.17}              &    {0.39}      &     {0.17}              \\
mSEM-UKF          &     {0.17}       &      {0.28}              &   {0.11}         &     {0.24}         &     {0.40}     &       {0.23}               \\
mSOR-UKF          &    {0.15}        &     {0.08}                &           {0.10} &      {0.07}             &   {0.36}       &       {0.06}            \\
\bottomrule
\end{tabular}
\label{Tab2}
\end{table*}

To test the comparative performance of the proposed method in practical settings, we carry out experimental campaigns at three sites for real-time indoor positioning using Ultra-wide band (UWB) devices. We use MDEK1001 Development Kit, by Qorvo, which includes 12 UWB units based on the DWM1001 module. {The module’s on-board firmware drives the built-in UWB transceiver to form a network of anchor nodes and perform the two-way ranging exchanges with the tag nodes which enables each tag to compute its relative location to the anchors. For detailed information regarding the kit functionality readers can consult its freely available documentation}. Figs.~\ref{Exp_sc_results_1}-\ref{Exp_sc_results_3} show the experimental scenarios at three different locations including a corridor in the School of Science and Engineering (SSE) building, a corridor in the Academic Block (AB) and the entrance of the SSE building at the Lahore University of Management Sciences. For each of the experimental scenario, 1 unit is configured as a tag and rest of the 11 units are set in the anchor mode. In each scenario, all the 11 anchors are installed at predefined locations. Moreover, the tag traverses through a predetermined path whose step coordinates are known. The kit is used in its default configuration mode in which at a given time a maximum of 4 range readings are obtained from the closest anchors. The tag is attached to a laptop which logs the range data at each step traversed. The range data is obtained at 5Hz which is subsequently averaged to provide the final sensor readings at each step. The datasets generated from the experimental campaigns are available openly: \url{https://github.com/chughtaiah/UWB_Data}.
\subsubsection*{Sources of outliers in sensors' data}
There are two major sources of outliers in the range data. 
\begin{figure}[h!]
\centering
\includegraphics[width=\columnwidth]{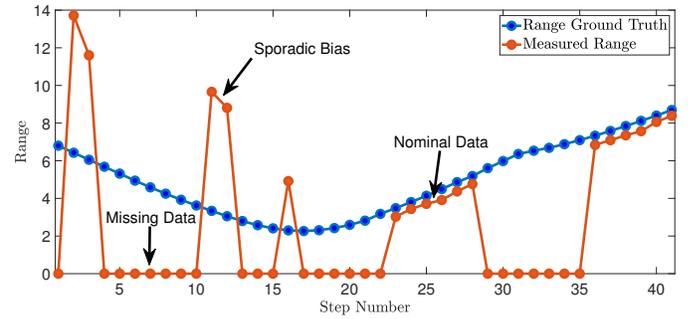}
\caption{Example of range data corruption obtained from a UWB sensor during experimentation}
\label{data_outliers}
\end{figure}
Firstly, there are missing observations in the data, since at most 4 out of 11 anchors provide data at a given time. The missing measurements are outliers in the data and contain no useful information. Secondly, UWB range data suffers from bias when the corresponding transceivers face physical obstruction during transmission known as the nonline-of-sight (NLoS) condition. Fig. \ref{data_outliers} shows example of how the range data obtained from a UWB sensor node during the experimental campaign is corrupted by the two types of outliers.

\subsubsection*{Performance results}
We consider random walk as the state mobility dynamic model which has historically been used for inference in different applications including mobile nodes in wireless sensor networks \cite{camp2002survey}. Keeping the 2D position of the target as our quantity of interest, the state vector $\mathbf{x}_k= [a_k,b_k]^\text{T}$ evolves with $\textbf{f}(.)=\mathbf{I}$ in \eqref{eqn_model_1}. Moreover, the nominal measurement model, $\textbf{h}(.)$ in \eqref{eqn_model_2}, has a functional form of range data as in \eqref{eqn_res3b} including the term for the $z$-axis for the locations of the tag and the anchors. In addition, $\mathbf{Q}_{k-\text{1}}$ and $\mathbf{R}_{k}$ are diagonal matrices with entries as 0.1. For each case, we set $\mathbf{x}_0= [0,0]^\text{T}$ {and initialize the filters randomly with ${\mathbf{m}}^+_{0} \sim \mathcal{N}(\mathbf{x}_0,\mathbf{P}^+_{0}$) where  $\mathbf{P}^+_{0}= 0.5 \mathbf{I}$. We} carry out 100 independent MC runs, keeping all other {applicable} parameters for the methods under consideration same as the {base parameters}.

\begin{figure}[!h]
\centering
\includegraphics[width=\columnwidth]{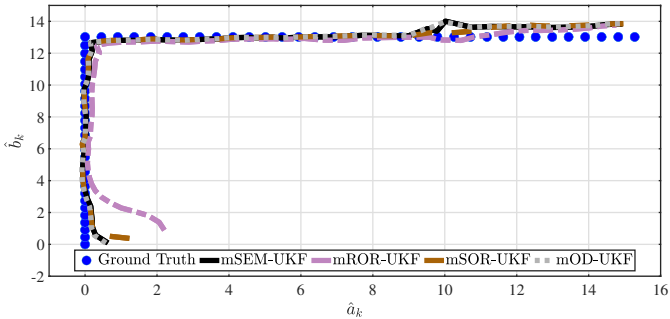}
\caption{Tracking performance for Case 1}
\label{Exp_results_1}
\end{figure}

\begin{figure}[!h]
\centering
\includegraphics[width=\columnwidth]{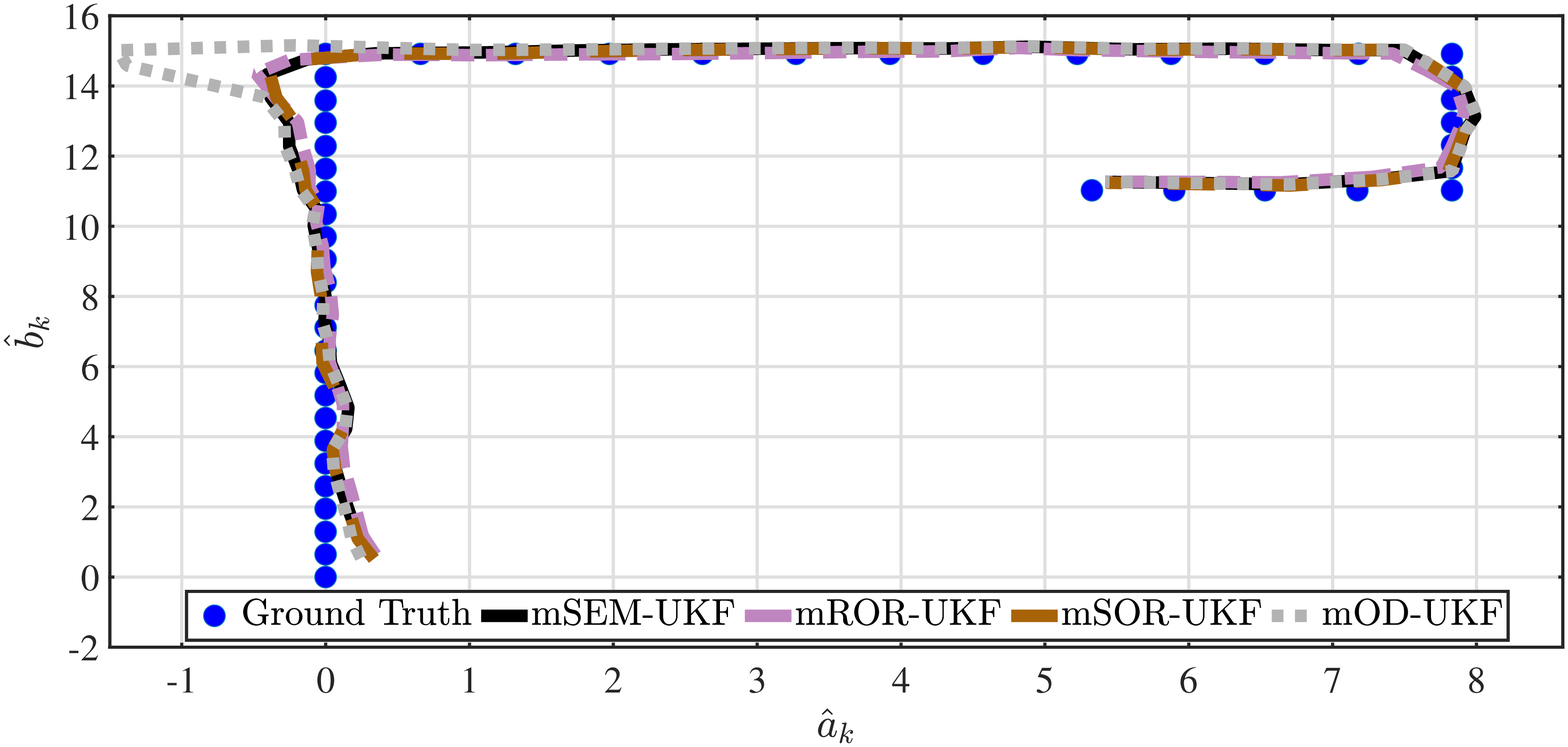}
\caption{Tracking performance for Case 2}
\label{Exp_results_2}
\end{figure}

\begin{figure}[!h]
\centering
\includegraphics[width=\columnwidth]{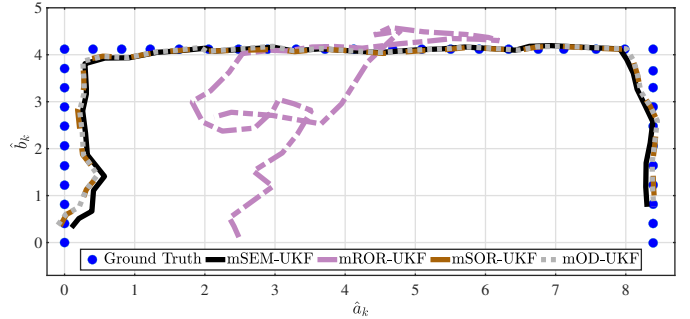}
\caption{Tracking performance for Case 3}
\label{Exp_results_3}
\end{figure}

We consider only the methods dealing outliers selectively with least computational complexity. Figs. \ref{Exp_results_1}-\ref{Exp_results_3} show the tracking performance of these filters. We observe that for cases 1 and 2 the methods exhibit similar tracking performances since the measurement mostly have missing observations as outliers for these cases. For case 3, the since data of some anchors is unusually contaminated with sporadic bias, in addition to the missing observations, the tracking error is larger in this case. In fact, tracking performance of mROR-UKF severely degrades for the third case due to unusual data corruption. 

Table \ref{Tab2} summarizes the performance evaluation results of the considered algorithms in the experimental settings in terms of RMSE and average computational run time for each case. We notice that the achieved RMSE in each case, normally remaining in the sub-meter range, is comparative for each algorithm. There is an exception for Scenario 3 where mROR-UKF exhibits larger errors due to more NLoS conditions resulting in occurrence of more bias in data. The proposed mSOR-UKF leads in terms of RMSE error and more importantly has the least processing overhead indicating its practical usefulness in real-world scenarios.


\section{Conclusion}\label{Concl}
Considering scenarios where a set of independent sensors provide observations for dynamical systems, we propose to model the outliers independently in each dimension for such cases. We devise an outlier-robust filter, resulting in selective rejection of corrupted measurements during inference. In addition, we propose modifications to the existing tractable learning-based outlier robust filters to deal outliers selectively. Also a modification to the proposed method is presented which yield lower computational complexity. Simulations reveal that the techniques which treat outliers selectively exhibit comparative estimation quality which is better as compared to the other methods. Moreover, the theoretical computational overhead is verified during simulations. Lastly, experimentation in various indoor localization scenarios, using UWB modules, suggests the practical efficacy of the proposed method. The gains obtained in terms of computational overhead can be critical where 1) the data is obtained from a large number of sensors and data acquisition rates are very high 2) the processing power is limited 3) energy savings are of prime concern for example in battery operated devices. 

\section*{Acknowledgement}
The authors thank Mr. Arslan Majal (serving as a research assistant in the Smart Data Systems \& Applications Laboratory LUMS) for his efforts in the experimental campaign. 

\bibliographystyle{IEEEtran}
\bibliography{biblio}

\end{document}